\documentstyle[preprint,aps]{revtex}
\input epsf

\newtheorem{rule1}{RULE}

\begin{document}
\draft
\preprint{cond-mat/9405043}
\title{Dynamical $\bbox{T=0}$ correlations of the $\bbox{S=1/2}$ 1D
Heisenberg Anti-Ferromagnet with $\bbox{\frac{1}{r^{2}}}$ exchange in
a
magnetic field}

\author{J.C. Talstra and F.D.M. Haldane}
\address{Joseph Henry Laboratories, Princeton University, Princeton,
NJ 08544}
\date{\today}

\maketitle
\begin{abstract}

We present a new selection rule for matrix elements of local spin
operators in
the $S=1/2$ ``Haldane-Shastry'' model.  Based on this rule we extend
a recent
exact calculation \cite{H93} of the ground-state dynamical spin
correlation
function $S^{ab}(n,t)$ = $\langle 0 | S^a(n,t)S^b(0,0)| 0 \rangle$
and its
Fourier-transform $S^{ab}(Q,E)$ of this model to a finite magnetic
field.  In
zero field, only {\it two-spinon} excitations contribute to the
spectral
function; in the (positively) partially-spin-polarized case, there
are two
types of elementary excitations:  {\it spinons} ($\Delta S^z = \pm
1/2$) and
{\it magnons} ($\Delta S^z = -1 $).  The magnons are divided into
left- or
right-moving branches.  The only classes of excited states
contributing to the
spectral functions are:  (I) two spinons, (II) two spinons + one
magnon, (IIIa)
two spinons + two magnons (moving in opposite directions), (IIIb) one
magnon.
The contributions to the various correlations are:  $S^{-+}$:  (I);
$S^{zz}$:
(I)+(II); $S^{+-}$:  (I)+(II)+(III).  In the zero-field limit there
are no
magnons, while in the fully-polarized case, there are no spinons.  We
discuss
the relation of the spectral functions to correlations of the
Calogero-Sutherland model at coupling $\lambda = 2$.
\end{abstract}
\pacs{75.10.Jm,71.10.+x}

%\narrowtext
\section{Introduction}
\label{intro}

In 1988 Haldane and Shastry \cite{H88,S88} independently introduced a
$S=\frac{1}{2}$ 1D spin model on $N$ sites with an exchange
interaction that
falls off inversely proportional to the distance between the spins.
In the
past few years this model has proven to be solvable to a remarkable
extent
\cite{H91,HHTPB92,H94}.  The simple structure of this model even
allowed the
authors of ref.  \cite{H93} to compute the zero magnetic field
dynamical
structure function (DSF) at zero temperature:
\begin{equation}
\label{strucdef}
\langle {\rm GS}|S_{m}^{a}(t)S_{n}^{b}(t')|{\rm GS}\rangle =
S^{ab}(m-n,t-t'),
\end{equation}
($a,b=x,y,z$).  In this paper we extend these results to a dynamical
groundstate correlation function in a {\em nonzero} magnetic field.
Although a
closed expression is not available, we are able to identify relevant
excitations that contribute to these functions.  More specifically:
when we
expand the expression (\ref{strucdef}) for the dynamical structure
functions in
a basis of eigenstates of the Hamiltonian $ \{ \left| \nu\right.
\rangle \}$:
\begin{eqnarray}
\label{strucexpansion}
\lefteqn{\langle {\rm GS}|S_{m}^{a}(m,t)S_{n}^{b}(n,t')|{\rm
GS}\rangle = }
\nonumber\\
& &\sum_{\nu} \langle\left.  {\rm GS}\right|
S^{a}_{m}(t)\left|\nu\right.
\rangle\langle\left.  \nu\right| S^{b}_{n}(t')\left|{\rm GS}\right.
\rangle\,
e^{-\frac{i}{\hbar}(t'-t)(E_{\nu}-E_{0})},
\end{eqnarray}
the set of intermediate states $\left|\nu\right.  \rangle$ that
contribute to
the sum is finite and contains only states that have very small
numbers of
elementary excitation added to the groundstate, viz.\ two ``spinons''
and up to
two ``magnons''.  In zero magnetic field the magnon excitations are
absent (see
\cite{H93}), whereas in very strong magnetic fields (such that all
spins are
fully polarized in the groundstate) only one magnon participates.
The set of
intermediate states is small as a consequence of a new selection rule
for
matrix elements of the local spin operators between eigenstates of
this model.
We will present it below.

A more traditional playground for 1D spin chains is the Heisenberg
model with
nearest neighbor exchange (NNE)\cite{B31}.  It shares its low energy
properties
with the Haldane-Shastry model (HSM).  However, for the NNE-model the
number of
excited states contributing to its dynamic structure functions is not
bounded;
consequently these functions are still unknown in both zero and
nonzero
magnetic field.  It seems again that the characterization of the HSM
as an
``ideal spinon gas'' \cite{H91} allows one to push forward our
understanding of
these 1D spin chains much further.

For the remaining part of this presentation, we will continue with a
brief
reiteration of the relevant properties of the Haldane-Shastry model
(HSM) in
Section~\ref{yangian}.  Section~\ref{selecrule} describes the
selection rule.
We then use this rule to identify the contributions of intermediate
states to
the different DSFs in Sections~\ref{smpsec} to \ref{spmsec}.  Finally
in
Section~\ref{mulsec} we conclude with a comparison to earlier DSF
calculations
for the NNE-model.

\section{Yangian Symmetries}
\label{yangian}

The Hamiltonian of the HSM in a magnetic field on $N$ sites is given
by:
\begin{equation}
\label{hamdef}
H=\frac{\pi v_s}{N^2}\sum_{i\neq j}d^{-2}(i-j) (P_{ij}-1) +
h\sum_{i}S^{z}_{i},
\end{equation}
where $d(n)=\sin(\frac{n\pi}{N})$ and $P_{ij}$ is the operator that
permutes
two spins on sites $i$ and $j$.  As described in
\cite{H88,HHTPB92,H94}, in
zero magnetic field the spectrum of the HSM consists of large
degenerate
multiplets.  Responsible for these degeneracies is a symmetry algebra
of
(\ref{hamdef}) identified in \cite{HHTPB92} as the Yangian $Y(sl(2))$
which is
generated by the following two vector operators:

\begin{eqnarray}
{\bf J}_{0} & = & \sum_{i=1}^{N} {\bf S}_{i} \nonumber\\
{\bf J}_{1} & = & \sum_{i<j} w_{ij} {\bf S}_{i}\times {\bf S}_{j},
\end{eqnarray}
where $w_{ij} = \frac{z_i + z_j}{z_i - z_j}$ and the $\{ z_i\},\,
i=1\ldots N$
are $N$ equally spaced points on the unit circle.  Both operators
commute with
the Hamiltonian.  Every degenerate multiplet forms a representation
of this
algebra and is characterized by a so-called (Yangian) highest weight
state
(YHWS).  That state is the only member of that multiplet which is
annihilated
by $J^{+}_{0}$ and $J^{+}_{1}$.  The rest of its multiplet is
generated when we
act upon the heighest weight state with the other members of the
Yangian
algebra.  Every Yangian multiplet has an accompanying {\em Drinfeld
polynomial}
$P(u)$ of order $\leq N$ in $u$.  The roots of this polynomial (which
can be
chosen to be integers or half-integers in the range $0\ldots N$)
correspond to
the elementary excitations of this model:  spinons.  A Drinfeld
polynomial can
be represented pictorially as a sequence $\Gamma$ of $N-1$ zeroes or
ones, a
generalized occupation number configuration.  We tag on two zeroes in
positions
0 and $N$.  As elucidated in \cite{H94} two ones have to be separated
by at
least one 0.  The roots of the Drinfeld polynomial are then located
between two
consecutive zeroes.  The locations of the ones, $\{m_i\}$, are the
so-called
{\em rapidities}.  There are $M=\frac{N-N_{sp}}{2}$ of them, where
$N_{sp}$ is
the number of spinons.  To clarify this with an example for $N=10$
sites, let
us consider the multiplet characterized by the sequence $101000100$.
It has
three rapidities $m_1 = 1, m_2 = 3, m_3 = 7$ and four spinons causing
roots in
the Drinfeld polynomial at $u=4 \frac{1}{2},\:  5 \frac{1}{2},\:  8
\frac{1}{2}$ and $9 \frac{1}{2}$.

Following \cite{H91}, a sequence of $m+1$ zeroes (between ones) is to
be
interpreted as a single orbital filled with $m$ spinons in a
symmetric state.
Since spinons have spin $1/2$ this implies that the orbital has total
spin
$\frac{m}{2}$ (more formally, this sequence gives an ``$m$-string''
of roots of
the Drinfeld polynomial, which, from the representation theory of the
Yangian
algebra \cite{C91}, constitutes a spin-$\frac{m}{2}$ factor in the
representation).  In the case of the above example multiplet that
means that
the multiplet has total spin content $1\otimes 1 = 2\oplus 1\oplus 0$
and
contains $3\times 3 = 9 $ states.  With $M$ ones there are therefore
$M+1$
orbitals---two of which are empty in our example.  Notice that the
maximum
possible $S^{z}$ and $S^{tot}$ a state in the multiplet can have is
$N_{sp}/2$.
That state (the YHWS) has all its spinons polarized up.

These spinons have a semionic character as is evidenced by the fact
that adding
two spinons (and we always have to add two at a time to avoid getting
a
sequence with consecutive ones) to a sequence reduces the number of
available
orbitals $M+1=\left(\frac{N-N_{sp}}{2}+1\right)$ by 1, as opposed to
0 for
bosons and 2 for fermions.  All the states in a multiplet
characterized by a
sequence $\{m_i\}$ have the same energy and momentum given by:
\begin{eqnarray}
P &= &\sum_{i=1}^{M}m_i\; {\rm mod}\:  N,\;\; {\rm in\; units\; of}\;
\frac{2\pi}{N},
\nonumber\\
\label{magnondisprel}
E & = &\sum_{i=1}^{M}\left(\frac{2\pi v_s}{N^2}\right) m_i(m_i-N).
\end{eqnarray}
We can also express these quantities in terms of spinon-variables.
If we label
the $M+1$ orbitals from right to left by spinon momenta $-k_0\leq
k\leq k_0 =
\frac{2\pi}{N} \frac{M}{2}$ spaced by $\frac{2\pi}{N}$ we get:

\begin{eqnarray}
P & = & \sum_{k}k n_{k\sigma} + Nk_0 \;{\rm mod}\; 2\pi ,\nonumber \\
\label{spinondisprel}
E & =&\sum_{k\sigma} \epsilon(k)n_{k\sigma} +
\frac{1}{N}\sum_{kk',\sigma\sigma'} V(k-k') +\; E(M,N),
\end{eqnarray}
where $\epsilon(k) = \frac{v_s}{\pi}(k_{0}^{2}-k^2)$,
$V(k)=v_s(k_0-|k|)$ and
$n_{k\sigma}$ is the number of spinons in the orbital with momentum
$k$ and
spin $\sigma$.  $E(M,N)$ only depends on the total number of sites
and spinons.
We also recognize $v_s$ as the spinon velocity $\frac{
d\epsilon(k)}{dk}$ at
the z\^{o}ne boundary, $k_0$, in the groundstate.

The action of the Yangian algebra within a multiplet of states is to
rotate the
spinons individually (rather than all of them through the global
$SU(2)$ spin
operators).  For this reason $J_{0}^{+}$ and $J_{1}^{+}$ annihilate a
YHWS
since it has all its spinons fully polarized.  Therefore the YHWS
have also
been dubbed Fully Polarized Spinon Gas States (FPSG) \cite{H91}.  In
a local
spin basis $ \{ \left| \{n_1,\ldots ,n_M\} \right.  \rangle \}$ where
the $\{
n_i\}$ are the locations of the down spins, the wavefunctions of
these FPSG
states $\Psi(n_1,\ldots,n_M)=\psi(z_{n_1},\ldots,z_{n_M})$ are
polynomials in
the $\{z_{n_i}\}$ of degree $<N$.  They can be written as $\chi
\Psi_0$, where
$\Psi_0$ is the $h=0$ groundstate Jastrow wavefunction, and $\chi$ is
a
polynomial known in the mathematical literature as a Jack polynomial.
Algorithms for their construction exist \cite{S72}.

To get a more physical idea of spinon states consider the following
wavefunctions in the same basis \cite{H94}:

\begin{eqnarray}
\lefteqn{\Psi(n_1,\ldots,n_M|\alpha _1,\ldots,\alpha _{N_{sp}})
=}\nonumber\\
&& \prod_{i<j}\left(z_{n_i}-z_{n_j}\right)^2 \prod_{i=1}^{M}z_{n_i}
\prod_{i=1}^{M}
\prod_{j=1}^{N_{sp}}\left(z_{n_i}-z_{\alpha_j}\right).
\label{locspinwf}
\end{eqnarray}
The $\{\alpha_i\}$ are the locations of {\em localized} spinons that
can point
$\uparrow$ or $\downarrow$; the $\{n_i\}$ are the positions of down
spins
(other than those of possible localized spinons pointing down).
Notice that
the wavefunction prevents the $\{n_i\}$ from coinciding with the
spinon sites.
We call the complement of the set of spinon sites the {\em
condensate}.  It is
a singlet under the action of total spin.  Furthermore for $N_{sp}=0$
eq.\
(\ref{locspinwf}) represents the exact groundstate wavefunction for
$h=0$.  The
usefulness of these states is limited by the fact that they are not
mutually
orthogonal and worse, overcomplete.  However, based on numerical
evidence for
up to 12 spinons, it is clear that the space spanned by states
(\ref{locspinwf}) with a fixed number $N_{sp}$ of {\em localized}
spinons
contains only eigenstates of the Hamiltonian belonging to Yangian
multiplets
with $N_{sp}$ or less spinons.

The subspace of states that have a fixed number of $N_{sp}$ localized
spinons,
all polarized, has the pleasant property that it only contains
eigenstates of
$H$ with {\em precisely} $N_{sp}$ spinons.  This is clear from the
fact that
these loc.\ spinon wavefunctions---although not eigenstates of the
Hamiltonian---are annihilated by both $J^{+}_{0}$ and $J^{+}_{1}$
(see Appendix
A).  This is consistent with the fact that fully polarized spinon
eigenstates
are supposed to be of a polynomial form with degree $<N$ in the
$\{z_{n_i}\}$\cite{H91}, just like wavefunctions (\ref{locspinwf}).
In that
same article it is also shown that these fully polarized localized
spinon
states are {\em complete} as well and span {\em all} YHWS.  \\
\vspace*{2ex}

In a nonzero magnetic field the term:  $h\sum_{i} S^{z}_{i}$ in the
Hamiltonian (\ref{hamdef}) will give a Zeeman splitting within the
Yangian
multiplets, although its members remain eigenstates of $H$.  As a
consequence,
for increasing magnetic field, the groundstate will contain more and
more
(fully polarized) spinons.  In the thermodynamic limit their number
is given
by:

\begin{equation}
\label{hmag}
\frac{N_{sp}}{N} = 2\sigma =  1- \sqrt{1-\frac{h}{h_c}}\; ; \; \;
h_c=\frac{\pi}{2}v_s,
\end{equation}
where $\sigma$ denotes the groundstate magnetization.
For $h\geq h_c$ the groundstate is completely ferromagnetic.

The occupation sequence characterizing the Yangian multiplet that
contains the
groundstate, will have the spinons ``condense'' into the left- and
rightmost
orbital (in equal numbers),  in accordance with the spinon dispersion
(\ref{spinondisprel}) relation which assumes a minimum at $\pm
k_{0}$.  So for
a typical magnetic field below the critical value, the groundstate
would be
the YHWS of a Yangian multiplet that is described by a sequence like
000010101010000.  For higher fields the 1010101 pattern shrinks as
more
spinons go into left and right orbitals.  These wavefunctions happen
to be
known analytically \cite{H88}:

\begin{equation}
\label{hgswf}
\Psi(n_1,\ldots,n_M) = \prod_{i<j}\left( z_{n_i}-z_{n_j}\right)^2
\prod_{i=1}^{M}(z_{n_i})^{\frac{N}{2}-M+1}.
\end{equation}
\section{Structure Functions and the Selection Rule}
\label{selecrule}

The dynamic structure function $S^{ab}(m-n,t-t') = \langle\left.
{\rm
GS}\right| S^{a}(m,t) S^{b}(n,t') \left|{\rm GS}\right.  \rangle$
with
$S^{a}(m,t)$ = $\exp(-itH) S^{a}_{m} \exp(itH)$ measures the response
of the
system to excitations created by flipping/imposing a certain spin on
a
particular site $n$ in state $\left|{\rm GS}\right.  \rangle$ at time
$t'$ and
measuring its effect at time $t$ on site $m$.  Obviously, since $H$
conserves $
S^{z}$, only $S^{-+}$, $S^{zz}$ and $S^{+-}$ are nonzero.  At zero
temperature
the Fourier transform $S(Q,E)$ of its expansion
(\ref{strucexpansion}) in a
basis of eigenstates $\{\left|\nu\right.  \rangle\}$ with energy
$E_{\nu}$ and
momentum $p_{\nu}$ looks like:

\begin{eqnarray} S(Q,E)&
=&\sum_{\nu} M_{\nu}^{a} \delta(E-(E_{\nu}-E_0))\,
\delta(Q-(p_{\nu}-p_0))
\nonumber\\
M_{\nu}^{a} &= &2\pi \left|\langle\left.  \nu\right|
S^{a}(Q)\left|{\rm
GS}\right.  \rangle\right|^{2},
\end{eqnarray}
with $S^{a}(Q)=\frac{1}{\sqrt{N}}\sum_{n=1}^{N}\exp(-inQ)S^{a}_{n}$,
and
$\left|{\rm GS}\right.  \rangle$ the groundstate of the model.  I.e.
the
support of $S^{ab}(Q,E)$ in the $(Q,E)$ plane is zero except when
$(Q,E)$
corresponds to the excitation-energy and -momentum of a state
contained in
$S^{a}(Q) \left|{\rm GS}\right.  \rangle$\cite{M81}.

{}From numerical evidence up to $N=16$ sites it has become clear that
there are
only an unexpectedly small number of nonzero matrix elements
$M^{a}_{n}$.  To
resolve parity and other accidental degeneracies between the Yangian
multiplets
we split these degeneracies by actually diagonalizing $H + \lambda
H_3$, where
$H_3$ is the second integral of the motion for this model as
presented in
\cite{HHTPB92,I90}.  The eigenvalues of this operator allowed the
Yangian
occupations sequences to be unambiguously identified.  States in the
multiplets
are partially resolved by fixing $ S^{z}$ and $S^{tot}$ (a unique
resolution of
states would be obtained by adding another term $\mu {\bf
J_0}\cdot{\bf J_1}$
to the Hamiltonian.  This would correspond to a basis of states
within the
Yangian multiplet obtained through the Algebraic Bethe Ansatz
\cite{F82}).

Now let us denote the eigenstates of this model as $\left|\Gamma
,\mu\right.
\rangle$ where $\Gamma$ labels a Yangian multiplet through an
occupation
sequence and $\mu$ labels the state within the multiplet.
Furthermore define
$M^{\Gamma\Gamma '}(S^{a}_{i})_{\mu\mu '}\equiv\langle\left.
\Gamma\mu\right|
S^{a}_{i}\left|\Gamma\mu '\right.  \rangle$.  Then the observation
made above
implies that the matrix $\underline{\underline{\bf M}}^{\Gamma\Gamma
'}(S^{a}_{i})$ vanishes if the occupation sequences $\Gamma$ and
$\Gamma '$
differ ``too much'' in a sense made precise below.  This situation is
analogous
to an {\em ideal} gas, where if $\hat{O}$ is a one body operator:
$\langle\left.  \alpha\right| \hat{O}\left|\beta\right.  \rangle = 0$
if the
occupation number configurations of $\left|\alpha\right.  \rangle$
and
$\left|\beta\right.  \rangle$ differ on more than one orbital.  E.g.\
$\hat{O}=\rho(x)=\sum_{kk'}e^{i(k-k')x}c_{k}^{\dagger}c_{k'}$ can add
or take
out a single particle from an orbital in an ideal gas, but in an {\em
interacting} gas it could add unlimited numbers of particle/hole
pairs.

\begin{figure}
\vspace{.2in}
\centerline{\epsfbox[53 705 298 739]{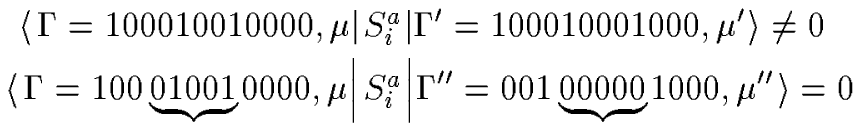}}
\vspace{.4in}
\caption{According to the selection rule, there will be states $\mu$
in
multiplet $\Gamma$ that are connected to others $\mu'$ in $\Gamma '$
through a
local spinon operator.  In $\Gamma$ and $\Gamma ''$ there aren't,
e.g.\ since
$|\pi (\Gamma ,4,8)-\pi(\Gamma' ,4,8)|=2-0=2>1$}
%\Gamma   = 100010010000
%\Gamma'  = 100010001000
%\Gamma'' = 001000001000
\label{selrulepic}
\end{figure}

However if $\Gamma$ and $\Gamma '$ do {\em not} differ too much,
according to
the rule there will always be a pair of states $\mu$ and $\mu '$ in
either
multiplet for which the matrix element is nonzero.
\begin{rule1}
If $\pi (\Gamma,m,n)$ is the total number of ones in positions $m$
through $n$
in Yangian occupation sequence $\Gamma$ then:
$\underline{\underline{\bf
M}}^{\Gamma\Gamma '}(S^{a}_{i})\neq\underline{\underline{\bf 0}}$
iff.
$|\pi(\Gamma ,m,n)-\pi(\Gamma ',m,n)|\leq 1$ for any $1\leq m<n\leq
N-1$.
\label{rule11}
\end{rule1}
The rule is illustrated in Fig.\ \ref{selrulepic}.  A general
consequence of
this rule is that when we choose $m=1$ and $n=N-1$, it follows that
the total
number of ones in a sequence can't change by more than one, i.e.\ the
total
number of spinons can only change by +2,0, or -2.  It is remarkable
that
according to the rule this also holds on any corresponding
subsequences of the
occupation number sequences.

The {\rm zero} magnetic field DSF has been computed in \cite{H93}.
The
particular structure function computed happened to be $S^{-+}(Q,E)$
(the others
are identical because of rotational invariance).  This function is
governed by
excitations present in $S^{+}_{i}\left|{\rm GS}\right.  \rangle$.
Since the
zero field groundstate contains no spinons the rule tells us that we
can only
expect excitations with 0 or 2 spinons.  As $ S^{+}_{i}\left|{\rm
GS}\right.
\rangle$ has $ S^{z} = +1$ the former is ruled out and in the states
in the
multiplets with 2 spinons, both must be polarized.  This was to be
expected
since we can expand $S^{+}_{i}\left|{\rm GS}\right.  \rangle$ in a
set of
localized spinon wavefunctions (\ref{locspinwf}) containing {\em two}
polarized
spinons:

\begin{eqnarray}
\lefteqn{\langle\left.  n_{1},\ldots,n_{M-1}
\right|S^{+}_{i}\left|{\rm
GS}\right.  \rangle =}\nonumber\\
&& \sum_{m=1}^{N-1}
\frac{2}{N}\frac{1-(-)^m}{1-z^{-m}} \Psi(n_1,\ldots,n_{M-1}|i,i+m).
\label{hnullspexp}
\end{eqnarray}
So one of the spinons seems to be sitting on the site on which
$S^{+}_{i}$
acted and the other is an even number of sites removed from it.

\begin{figure}
\centerline{\epsfbox[135 259 483 540]{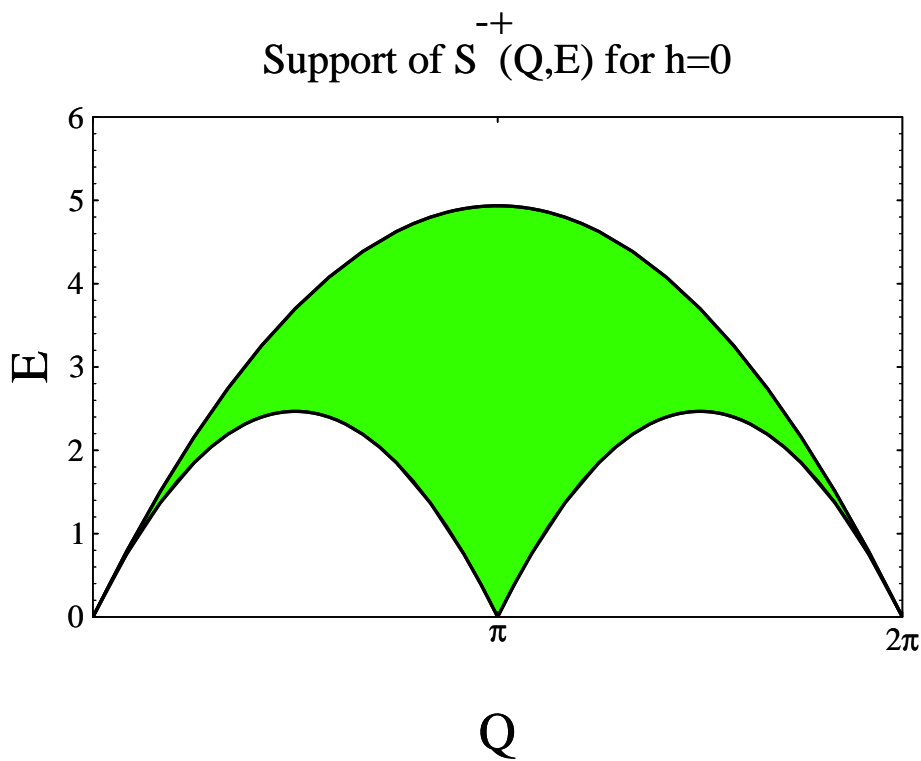}}
\vspace{.4in}
\caption{The shaded region show where $S^{-+}(Q,E)$ is nonzero for
$h=0$.  The
top boundary corresponds to excitations with two spinons that have
identical
momentum; on the bottom boundary one of the spinons has fixed
momentum $\pm
\pi$. $E$ is given in units of $v_s /\pi$.}
\label{smphne0}
\end{figure}

Since we know the spinon dispersion relation, we can demarcate the
support of
$S^{-+}(Q,E)$ in the $(Q,E)$-plane in Fig.\ \ref{smphne0}.  The main
steps of
the computation of the matrix elements $M_{\nu}^{+}$, i.e.\ the
weight of the
DSF at a point $(Q,E)$ on the plot, are the following:  since
$S^{+}_{i}\left|{\rm GS}\right.  \rangle$ only contains states with
two fully
polarized spinons, it must be built out of YHWS.  These wavefunctions
are
functionally identical to eigenfunctions of the Calogero-Sutherland
model at
coupling $\lambda =2$ of particles moving on a ring.  Since both
wavefunctions
are of a polynomial form with degree $<N$ the computation of a sum
over sites
is identical to taking an integral over the ring in the continuum
model.  The
action of $S^{+}_{i}$ in the spin chain is translated into a particle
destruction operator $\Psi(x,t)$ ($ S^{+}_{i}$ removes a down spin).
So the
$S^{+-}(Q,E)$ DSF reduces to the Greens function $\langle\left.  {\rm
GS}\right| \Psi^{\dagger}(x,t)\Psi(0,0)\left|{\rm GS}\right.
\rangle$ in the
Calogero-Sutherland model.  It can be computed in the thermodynamic
limit, in
which case it can be mapped unto a Gaussian hermitian matrix model
correlator.
The result is:

\begin{equation}
\label{psidagpsi}
S^{-+}(Q,E) = \frac{1}{8v_s}\left(\frac{(v_1-v_2)^2}{( v_s ^2
-v_{1}^{2})( v_s
^2-v_{2}^{2}) }\right)^{\frac{1}{2}},
\end{equation}
with $ Q= -\frac{\pi}{2 v_s} (v_1+v_2)$ and
$E=\epsilon(v_1)+\epsilon(v_2)$.
The DSF matrix element is parametrized by $v_1,v_2$, which are
quickly
identified with the velocities of the two spinons in the excited
state.  The
$\langle\Psi^\dagger\Psi\rangle$ Greens function has been obtained
recently at
finite $N$ as well \cite{Ha94}.

For $h\neq 0$ the three different structure functions $S^{-+}(Q,E)$,
$S^{zz}(Q,E)$ and $S^{+-}(Q,E)$ will not be equal, since $\left|{\rm
GS}\right.
\rangle$ is no longer a singlet.  In fact it has $ S^{z} =
S^{tot}=\frac{N_{sp}}{2}\equiv S_0$, where $N_{sp} = N_{sp} (h)$ is
given by
(\ref{hmag}).  This difference between the three correlation function
is also
expressed in two additional global $SU(2)$ selection rules\cite{M81},
which
rule out certain matrix elements based on the total spin $S^{tot}$
and $ S^{z}$
of the final state $\mu$ {\em inside} the Yangian multiplet $\Gamma$.

In the first place there is the Wigner-Eckart theorem for vector
operators such
as $S^{a}_{i}$, which tells us that in order for $\langle\left.
\Gamma
,\mu\right| S^{a}_{i}\left|{\rm GS}\right.  \rangle$ to be nonzero
the total
spin $S$ of $\left|\Gamma,\mu\right.  \rangle$ must satisfy $S_0 -1
\leq S\leq
S_0 +1$.  Secondly, for any state in a multiplet with $N_{sp}$
spinons, we have
$\frac{N_{sp}}{2}\geq S\geq S^{z}$--- where equality only holds for
the YHWS,
which has all its spinons $\uparrow$.  Now put this together with the
fact that
$ S^{+}_{i}$ raises $ S^{z}$ by +1, $ S^{-}_{i}$ lowers it by 1, and
$
S^{z}_{i}$ leaves it the same.  Classifying states according to their
$ S^{z}$
and $S^{tot}$ as types $(i)$ - $(vi)$, following M\"{u}ller et al.\
\cite{M81}
we find the following contributions:
\begin{description}
\item[\underline{$ S^{+}_{i}\left|{\rm GS}\right.  \rangle$}]
contains states
with $S=S_0+1$ and $\Delta N_{sp} =+2$ (type $(iii)$ ).
\item[\underline{$ S^{z}_{i}\left|{\rm GS}\right.  \rangle$}]
contains states
with $S=S_0+1$ and $\Delta N_{sp} =+2$ (type $(i)$ ) or $S=S_0$ and
$\Delta
N_{sp} =+2,0$ (type $(ii)^{a,b}$ ).
\item[\underline{$ S^{-}_{i}\left|{\rm GS}\right.  \rangle$}]
contains states
with $S=S_0+1$ and $\Delta N_{sp} =+2$ (type $(iv)$ ) or $S=S_0$ and
$\Delta
N_{sp} =+2,0$ (type $(v)^{a,b}$ ), or $S=S_0-1$ and $\Delta N_{sp}
=+2,0,-2$
(type $(vi)^{a,b,c}$ ).\\
\end{description}
Since we have an additional quantum number to label states:
$N_{sp}$, we added
Latin superscripts $a,b,c$ to the Roman numerals.  All 10
contributions are
summarized in table (\ref{sumtable}).

We will now investigate all three structure functions individually
following
these selection rules.

\section{${\bf S^{-+}(Q,E)}$}
\label{smpsec}

\vspace{2.5ex}
\underline{Type $(iii)$: $\Delta N_{sp} =+2$, $\Delta S=+1$.}

For the occupation sequence of the groundstate in a given magnetic
field (e.g.\
000010101010000) let us label the zeroes in the leftmost orbital as
the left
spinon condensate and the ones in the rightmost orbital as the right
spinon
condensate.  From table \ref{sumtable} we learn that action of $
S^{+}_{i}$ on
the groundstate only produces states with two more spinons, i.e.\ one
less 1.
This 1 has to come out of the center $\cdots 10101\cdots$ region.  We
can't
take more than a single 1 out of the center region---and stow it into
the left
or right spinon condensate---since this would imply a violation of
Rule
\ref{rule11} applied to the center region.  Taking out a 1 in the
center region
is equivalent to inserting 2 spinons there.  A typical nonzero matrix
element
would be $\langle\left.  0001001001000,\mu\right| S^{+}_{i} \left|
0001010101000\right.  \rangle$, with the two spinons residing in
orbital two
and three.

All this means is that we get a simple two spinon spectrum, as in the
zero
magnetic field case.  The only difference is that now the momenta of
the
spinons can only vary from $-k_0$ to $k_0$, where
$k_0=\frac{\pi}{4N}(N-N_{sp})$ decreases with increasing magnetic
field as
$N_{sp}=N_{sp}(h)$ according to eq.\ (\ref{hmag}).  The support of
$S^{-+}(Q,E)$ is essentially a squeezed version of Fig.\
\ref{smphne0}.

As for the weight associated with 2-spinon excitations:  the
calculation for
the zero magnetic field case (\ref{psidagpsi}) carries over without
problems.
The reason for this is twofold.  In the first place, the nonzero
magnetic
field groundstate wavefunction (\ref{hgswf}) is of the same
Jastrow-form as the
zero field one, with just an extra phase factor
$\prod_{i}z_{n_i}^{\frac{N}{2}-M}$ appended.  When we take the matrix
element,
the phase factors from ket and bra part cancel each other.  Secondly,
the
excited states are again of the YHWS type $(S=N_{sp})$ and have to be
polynomials.  The mapping onto a Calogero-Sutherland model matrix
element
remains therefore legitimate.

The contribution of just two-spinon YHWS excitations was to be
expected since
we can expand any fully polarized localized spinon wavefunction with
$N_{sp}$
spinons acted upon with $ S^{+}_{i}$ in terms of a set containing
$N_{sp}+2$
spinons:

\begin{eqnarray}
\lefteqn{
( S^{+}_{i}\Psi_{\alpha_i})( n_{1},\ldots,n_{M-1} ) =}\nonumber\\
&& \sum_{p\in V}\prod_{r\in V}\frac{(z_r-z_i)}{(z_r-z_p)} \Psi(
n_{1},\ldots,n_{M-1} |\alpha_1,\ldots,\alpha_{N_{sp}},i,p).
\label{spin2exp}
\end{eqnarray}
Here $V$ is a set of $M$ random sites on the circle excluding the
spinon sites
$\{\alpha_i\}$ (see Appendix B).  Eq.  (\ref{hnullspexp}) is a
special case of
this expansion with $V$ equal to the sites that are an even number of
steps
removed from the site on which the local spin operator acts.  We
could also
have realized that the number of spinons can't go up by more than 2
when we
consider that $ S^+$ and $J^{+}_{1}$ annihilate $ S^{+}_{i}\left|{\rm
GS}\right.  \rangle$, indicating that the latter must consist of
purely YHWS
{\em with} $ S^{z} = S_{0} +1$!

\section{${\bf S^{zz}(Q,E)}$}
\label{szzsec}

For the DSF $S^{zz}(Q,E)$ we find similar simple excitations that
contribute,
although at present not all resulting matrix elements can be
computed.  From
the combined selection rules we find three types of excitations:
$(i)$,
$(ii)^a$ and $(ii)^b$, in table \ref{sumtable}.  They all have in
common that
$\Delta N_{sp} =0$ or +2.  This isn't surprising:  let us consider
the state
$J^{+}_{1}\left( S^{z}_{i}\left|{\rm GS}\right.  \rangle\right)$.
This state
is annihilated by $ J^+_1$ and $ S^+$, so it must be YHWS (see
Appendix C).
Therefore, since the first action of $ J^+_1$ doesn't change the
number of
spinons, $ S^{z}_{i}\left|{\rm GS}\right.  \rangle$ must be a mixture
of states
that contain not more than $2S_0 +2$ spinons, where $2S_0$ is the
number of
spinons in the groundstate.

We will now discuss the individual types and where possible compute
the values
of the matrix elements.

\vspace{2.5ex}
\underline{Type $(i)$: $\Delta N_{sp} =+2$, $\Delta S=+1$.}

Having identical selection rules, these states sit in the exact same
Yangian-
and spin multiplets as the type $(iii)$ states, and therefore they
also contain
two additional spinons.  However their $ S^{z} =S^{tot} -1$, so they
are no
longer YHWS like the type $(iii)$ states.  Nevertheless they are
related by a
simple application of $ S^{-}$:

\begin{equation}
|{\scriptstyle \Gamma,\Delta N_{sp} =+2},\!
		\renewcommand{\arraystretch}{.5}
		\begin{array}{c}
		{\scriptscriptstyle S^{tot}= S_0 +1} \\
		{\scriptscriptstyle  S^{z} =S_0}
	        \end{array} \rangle
= {\scriptstyle \frac{1}{\sqrt{2(S_0 +1)} }}
S^{-}\left|\Gamma\right. \rangle,
\end{equation}
where $\left|\Gamma\right.  \rangle$ denotes the YHWS of the
multiplet with
occupation sequence $\Gamma$ of type $(iii)$.  This allows us to
reduce a type
$(i)$ matrix element to one that is a multiple of a type $(iii)$
given in eq.\
(\ref{psidagpsi}):
\begin{eqnarray}
&&\left|\langle{\scriptstyle \Gamma,\Delta N_{sp} =+2} ,\!
		\renewcommand{\arraystretch}{.5}
		\begin{array}{c}
		{\scriptscriptstyle S^{tot}= S_0 +1} \\
		{\scriptscriptstyle  S^{z} =S_0}
	        \end{array} |
S^{z}_{i}\left|{\rm GS}\right. \rangle\right|^2
\rule{10em}{0em}\nonumber\\
&&\rule{2em}{0em}= {\scriptstyle \frac{1}{2(S_0
+1)}}\left|\langle\left.
\Gamma\right | [ S^+, S^{z}_{i} ]\left|{\rm GS}\right.
\rangle\right|^2
\nonumber\\
&&\rule{2em}{0em}={\scriptstyle \frac{1}{2(S_0
+1)}}\left|\langle\left.
\Gamma\right| S^{+}_{i}\left|{\rm GS}\right.  \rangle\right|^2,
\label{psidagpsisz}
\end{eqnarray}
where we used that $ S^+$ annihilates the groundstate.

\vspace{2.5ex}
\underline{Type $(ii)^a$: $\Delta N_{sp} =+2$, $\Delta S=0$.}

This type of state is a member of the same kind of {\em Yangian}
multiplet as
type $(iii)$ and $(i)$---i.e.\ with two extra spinons---but it sits
in a {\em
spin} multiplet that does not contain the YHWS.  We expect the
associated
matrix element to be proportional to a type $(iii)$ matrix element as
well, but
we lack the necessary operator that would step us form this state to
the YHWS.
This operator would have to be some member of the Yangian algebra.

\vspace{2.5ex}
\underline{Type $(ii)^b$: $\Delta N_{sp} =0$, $\Delta S=0$.}

Since this type of state has $\Delta N_{sp} =0$ the number of ones in
its
occupation sequence must be identical to that in the groundstate.  As
with the
previous three types we can delete just a single 1 from the center,
leaving
behind two spinons pointing up.  This 1 then must go into either the
left or
the right spinon condensate.  Therefore a typical nonzero matrix
element would
be:  $\langle\left.00100\rule[-.25em]{.3ex}{1.1em}
1001001\rule[-.25em]{.3ex}{1.1em} 00000\right| S^{z}_{i}\left|
00000\rule[-.25em]{.3ex}{1.1em} 1010101\rule[-.25em]{.3ex}{1.1em}
00000\right.
\rangle$ where the $\rule[-.25em]{.3ex}{1.1em}$ just helps to draw
attention to
the center region.  Rule \ref{rule11} rules out any additional ones
leaving the
center region.  The additional 1 on the left or right can be
interpreted as a
{\em magnon} with $ S^{z} = -1$.  The limiting case where the magnon
``fuses''
with two spinons at a boundary between a condensate and the center
region gives
us the groundstate.

These states must be YHWS since they have $\frac{N_{sp}}{2} =S=
S^{z}$, like
$\left|{\rm GS}\right.  \rangle$.  This fact allows us to calculate
the
corresponding matrix elements.  Since both groundstate and excited
state are
YHWS, a mapping onto the Calogero-Sutherland model is valid.  In this
case we
need a groundstate density-density correlator $\langle\left.  {\rm
GS}\right|
\rho(x,t)\rho(x',t')\left|{\rm GS}\right.  \rangle$ since $
S^{z}_{i}$ measures
the presence or absence of a down-spin (i.e.\ a particle in the
CS-model).
This calculation has been done by Altshuler et al.\ \cite{A93}, in
the
thermodynamic limit, by studying the repulsion of energy levels in a
random
matrix model under a varying perturbation.  The energy levels are
identified
with the positions of the particles and the strength of the
perturbation
corresponds to imaginary time.  Their original expression depends on
three
parameters (called $\lambda,\lambda_1$ and $\lambda_2$), the latter
two of
which are compact, and the first one is unbounded.  This is precisely
what we
expect from our selection rule:  2 spinons restricted to the center
region with
momenta in the range $-k_0\ldots +k_0$ and a magnon that can go off
all the way
to the right or left (i.e.\ $\pm \infty$ in the thermodynamic limit).
 In terms
of the velocity, $v$, of the magnon---with dispersion relation
(\ref{magnondisprel})---and spinon velocities $v_1,v_2$ their result
is as
follows:

\begin{eqnarray} \lefteqn{\left|\langle\left.  v,v_1,v_2\right|
\rho(Q)\left|{\rm GS}\right.  \rangle\right|^2 =}\nonumber\\
&& \frac{(v- v_s )(v+ v_s )}{(v-v_1)^2(v-v_2)^2} \left[
(v-v_1)+(v-v_2)\right]^2 .
\label{rhorho}
\end{eqnarray}
For completeness we give the relation between the $v$'s and the
$\lambda$'s :

\begin{eqnarray}
\lambda & =& v/ v_s \nonumber\\
\lambda_1\lambda_2 & = &\mbox{$\frac{(v_1 +v_2)}{2 v_s }$}\nonumber\\
(1-\lambda_{1}^{2})(1-\lambda_{2}^{2}) & = & \mbox{$\left(\frac{(v_1
-v_2)}{2
v_s }\right)^2 $}.
\end{eqnarray}

\begin{figure}
\centerline{\epsfxsize=6.5in
\epsfbox[28 381 761 743]{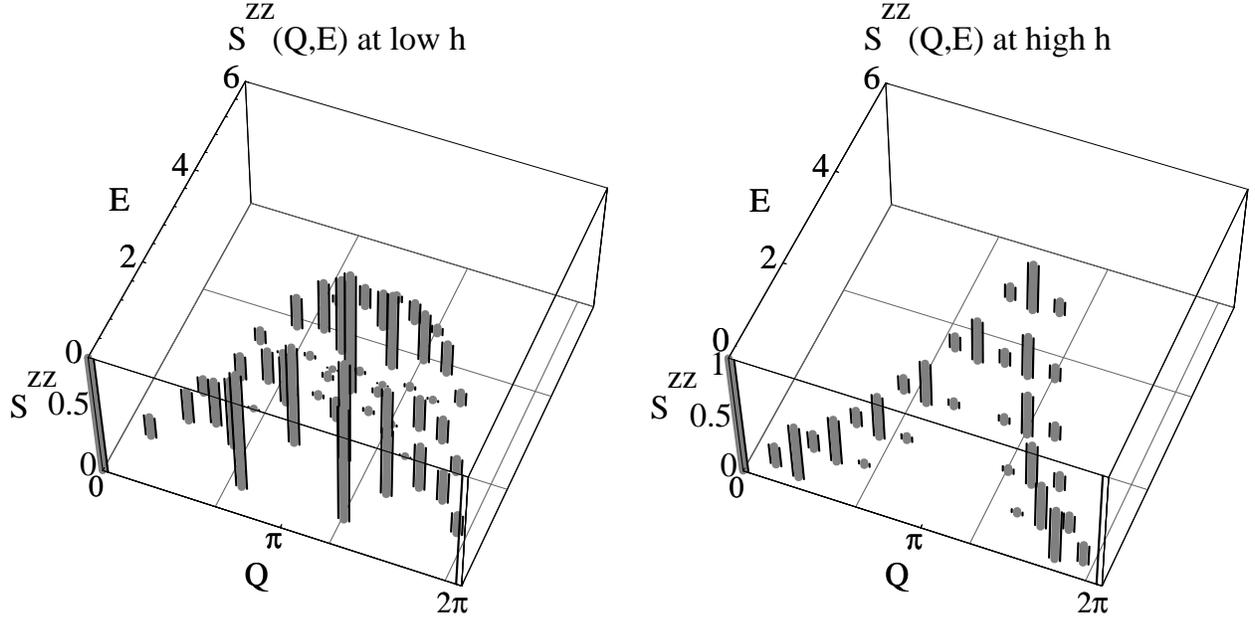}}
\vspace{.4in}
\caption{ $S^{zz}(Q,E)$ for small $h$ $(\sigma = \frac{S_z}{N}=.05)$
and large
$h$ $(\sigma=.4)$ on $N=14$ sites. $E$ is in units of $v_s /\pi$.}
\label{szzdata}
\end{figure}

\begin{figure}
\centerline{\epsfxsize=6.5in
\epsfbox[38 398 768 744]{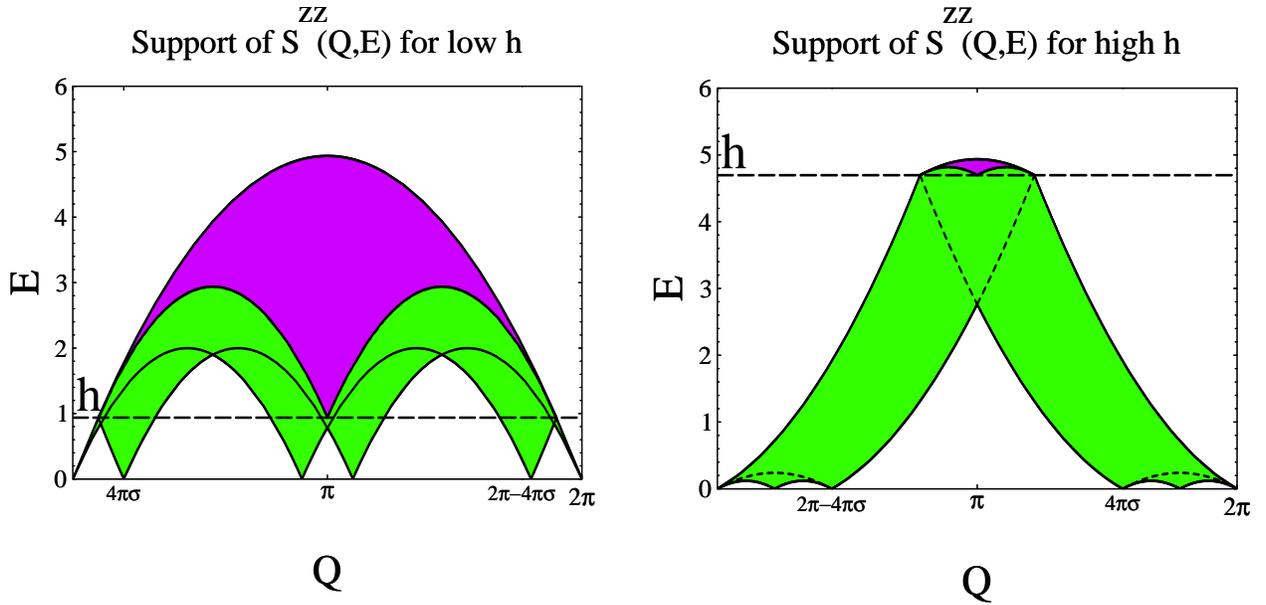}}
\vspace{.4in}
\caption{The regions where $S^{zz}(Q,E)$ is nonzero for low and high
$h$.  The
area shaded dark contains the contributions from the excitations with
$\Delta
N_{sp} = +2$ (types ($(i)$ and $(ii)^a$).  The excitations with
$\Delta
N_{sp}=0$---type $(ii)^b$---can also occupy the lightly shaded area.}
\label{szzsup}
\end{figure}

Numerical data for $S^{zz}(Q,E)$ can be found in Fig.\ \ref{szzdata}
at values
of $h$ close to 0 and $h_c$.  Fig.\ \ref{szzsup} shows the
corresponding
support of $S^{zz}(Q,E)$ in the $(Q,E)$ plane, as predicted from the
selection
rule, in the thermodynamic limit.  We notice that for finite-size
systems some
of the features near the lower boundary are absent.

\section{${\bf S^{+-}(Q,E)}$}
\label{spmsec}

Finally we discuss the $S^{+-}(Q,E)$ DSF which is governed by the
excitations
of types $(iv)$ - $(vi)^c$ that are present in $ S^{-}_{i}\left|{\rm
GS}\right.
\rangle$.

\vspace{2.5ex}
\underline{Type $(iv)$: $\Delta N_{sp} =+2$, $\Delta S=+1$.}

This type of state is very similar to types $(iii)$ and $(i)$, as a
matter of
fact they all reside in identical Yangian- and spin multiplets.
Therefore type
$(iv)$ states differ from the groundstate only by two extra spinons
in the
center region.  They are related to their YHWS $\left|\Gamma\right.
\rangle$
(of type $(iii)$) through:

\begin{equation}
| {\scriptstyle \Gamma,\Delta N_{sp} =+2},\!
\renewcommand{\arraystretch}{.5}
\begin{array}{c}
{\scriptscriptstyle S^{tot}= S_0 +1} \\
{\scriptscriptstyle S^{z} =S_0 -1}
\end{array}
\rangle ={\scriptstyle \frac{1}{\sqrt{(4S_0 +2)(2S_0 +2)}}}
(S^{-})^2\left|\Gamma\right.  \rangle.
\end{equation}
Analogous to the calculation for type $(i)$ states, a matrix element
of type
$(iv)$ can now easily be reduced to one involving type $(iii)$:

\begin{eqnarray}
&&\left|\langle{\scriptstyle \Gamma,\Delta N_{sp} =+2},\!
		\renewcommand{\arraystretch}{.5}
		\begin{array}{c}
		{\scriptscriptstyle S^{tot} = S_0 +1} \\
		{\scriptscriptstyle  S^{z} =S_0 -1}
	        \end{array} |
S^{z}_{i} \left|{\rm GS}\right. \rangle\right|^2
\rule{10em}{0em}\nonumber\\
&&\rule{2em}{0em}={\scriptstyle \frac{1}{(4S_0 +2)(2S_0 +2)}}\left|
\langle\left.  \Gamma\right| ( S^+)^2 S^{-}_{i}\left|{\rm GS}\right.
\rangle\right|^2 \nonumber \\
&&\rule{2em}{0em} = {\scriptstyle \frac{1}{(2S_0 +1)(S_0
+1)}}\left|\langle\left.  \Gamma\right| S^{+}_{i}\left|{\rm
GS}\right.
\rangle\right|^2.
\label{psidagpsismin}
\end{eqnarray}
The last matrix element in this equation has already been computed
for the
$S^{-+}(Q,E)$ DSF.  However the energy of the corresponding excited
state here
is shifted by $2h$ in comparison, because of the Zeeman term in the
Hamiltonian.

\vspace{2.5ex}
\underline{Types $(v)^a$ and $(vi)^a$:$\Delta N_{sp} =+2$, $\Delta
S=0, -1$.}

States of types $(v)^a$ and$(vi)^a$ contain a two-spinon excitation
as
well---like types $(iii)$, $(i)$ and $(iv)$.  However, since they
don't reside
in the spin multiplet of their YHWS ($\frac{N_{sp}}{2} >S$), the
associated
matrix elements are unknown.

\vspace{2.5ex}
\underline{Types $(v)^b$: $\Delta N_{sp} =0$, $\Delta S=0$.}

Type $(v)^b$ states are very similar to those of type $(ii)^b$, they
only
differ in $ S^{z}$ by -1.  Therefore both contain two excited spinons
and a
single left- or right moving magnon.  They are related to each other
by:

\begin{equation}
| {\scriptstyle \Gamma,\Delta N_{sp} =0},\!
		\renewcommand{\arraystretch}{.5}
		\begin{array}{c}
		{\scriptscriptstyle S^{tot}= S_0} \\
		{\scriptscriptstyle  S^{z} =S_0 -1}
	        \end{array} \rangle
={\scriptstyle \frac{1}{\sqrt{2S_0}}} S^{-}\left|\Gamma\right.
\rangle,
\end{equation}
and $\left|\Gamma\right.  \rangle$ is the type $(ii)^b$ YHWS of the
multiplet
with occupation sequence $\Gamma$.  We can now trivially relate the
matrix
elements of type $(ii)^b$ and $(v)^b$:

\begin{eqnarray}
&&\left|\langle{\scriptstyle \Gamma,\Delta N_{sp} =0},\!
		\renewcommand{\arraystretch}{.5}
		\begin{array}{c}
		{\scriptscriptstyle S^{tot}= S_0} \\
		{\scriptscriptstyle  S^{z} =S_0 -1}
	        \end{array} |
S^{-}_{i}\left|{\rm GS}\right. \rangle\right|^2
\rule{10em}{0em}\nonumber\\
&&\rule{2em}{0em}={\scriptstyle \frac{1}{2S_0}}\left|\langle\left.
\Gamma\right|[ S^+, S^{-}_{i}]\left|{\rm GS}\right.  \rangle\right|^2
\nonumber\\
&&\rule{2em}{0em}={\scriptstyle
\frac{2}{S_0}}\left|\langle\left.\Gamma\right|
S^{z}_{i}\left|{\rm GS}\right.  \rangle\right|^2 .
\label{rhorhosmin}
\end{eqnarray}
The last matrix element is listed in eq.\ (\ref{rhorho}).

\vspace{2.5ex}
\underline{Type $(vi)^b$: $\Delta N_{sp} =0$, $\Delta S=-1$.}

These states reside in the same Yangian multiplets as the previous
type however
they are not in the spin multiplets of the YHWS:
$\frac{N_{sp}}{2}>S=S_0 -1$.
Since we lack the operator that steps us up to the YHWS, we aren't
able to
compute the matrix element.

\vspace{2.5ex}
\underline{Type $(vi)^c$: $\Delta N_{sp} =-2$, $\Delta S=-1$.}

This last class of states has $\Delta N_{sp} =-2$ and therefore the
number of
ones in its occupation sequence goes up by one compared to the
groundstate.
The selection Rule \ref{rule11} only allows the extra 1 to go into
the left or
right spinon condensate.  As before we are also allowed to take a 1
out of the
center region and bring it into the left/right spinon condensate.
Notice
however that the rule forbids both of the ones to go into the {\em
same}
condensate:  one has to be left moving and the other must be right
moving.  The
result is an excited state with two magnons and two spinons.  A
typical nonzero
matrix element would be $\langle\left.
0100\rule[-.25em]{.3ex}{1.1em}
100101001\rule[-.25em]{.3ex}{1.1em} 0001\right| S^{-}_{i}\left|
0000\rule[-.25em]{.3ex}{1.1em} 101010101\rule[-.25em]{.3ex}{1.1em}
0000\right.
\rangle$.  The $\rule[-.25em]{.3ex}{1.1em}$ just helps to guide the
eye.  Also
present are YHWS of multiplets from the limiting cases where one of
the magnons
fuses with the two spinons in the center; this leaves a multiplet
with nothing
but one magnon; example:  $\langle\left.
0000\rule[-.25em]{.3ex}{1.1em}
101010101\rule[-.25em]{.3ex}{1.1em} 0010\right| S^{-}_{i}\left|
0000\rule[-.25em]{.3ex}{1.1em} 101010101\rule[-.25em]{.3ex}{1.1em}
0000\right.
\rangle$.  This single magnon excitation is familiar from the strong
field
regime.

Since type $(vi)^c$ states are YHWS ($\frac{N_{sp}}{2}=S$), as is the
groundstate, we can repeat the calculation of the matrix elements by
a mapping
onto the Calogero-Sutherland model.  Because $ S^{-}_{i}$ creates a
down spin,
it corresponds to a particle creation operator in the CS-model.  The
relevant
correlation function is therefore $\langle\left.  {\rm
GS}\right|\Psi(x,t)\Psi^{\dag}(x',t')\left|{\rm GS}\right.  \rangle$.
 As in
the $\langle\Psi^{\dag}\Psi\rangle$ case for type $(iii)$ states, a
further
mapping onto a Gaussian Hermitean matrix model allows one to
calculate the
Fourier transform of this correlation function \cite{Z93}.  The
result is
parametrized by four variables, two of which are compact:  $v_1,v_2$,
and two
are non-compact:$v,v'$:

\begin{eqnarray}
\lefteqn{\left|\langle\left.  v,v',v_1,v_2\right| S^{-}_{i}\left|{\rm
GS}\right.  \rangle\right|^2 \propto}\\
\label{psipsidag}
&& \frac{(v^2- v_s ^2)(v'^2- v_s ^2)|v_1-v_2|}{( v_s ^2-v_{1}^{2})(
v_s^2-v_{2}^{2}) (v^2-v_{1}^{2}) (v^2-v_{2}^{2}) (v'^2-v_{1}^{2})
(v'^2-v_{2}^{2})}.  \nonumber
\end{eqnarray}
Energy and momentum in terms of the $v$'s are given by
$Q=\frac{\pi}{v_s}(v+v'-\frac{1}{2} (v_1 + v_2))$ and
$E=\frac{\pi}{2v_s} (v^2
+ v'^2 - \frac{1}{2}(v_{1}^{2}+v_{2}^{2}) -v_{s}^{2})$.  It is
obvious that the
compact parameters are to be identified with the spinon velocities
and the
non-compact ones with the magnon velocities.

\begin{figure}
\centerline{\epsfxsize=6.5in\epsfbox[38 390 769 743]{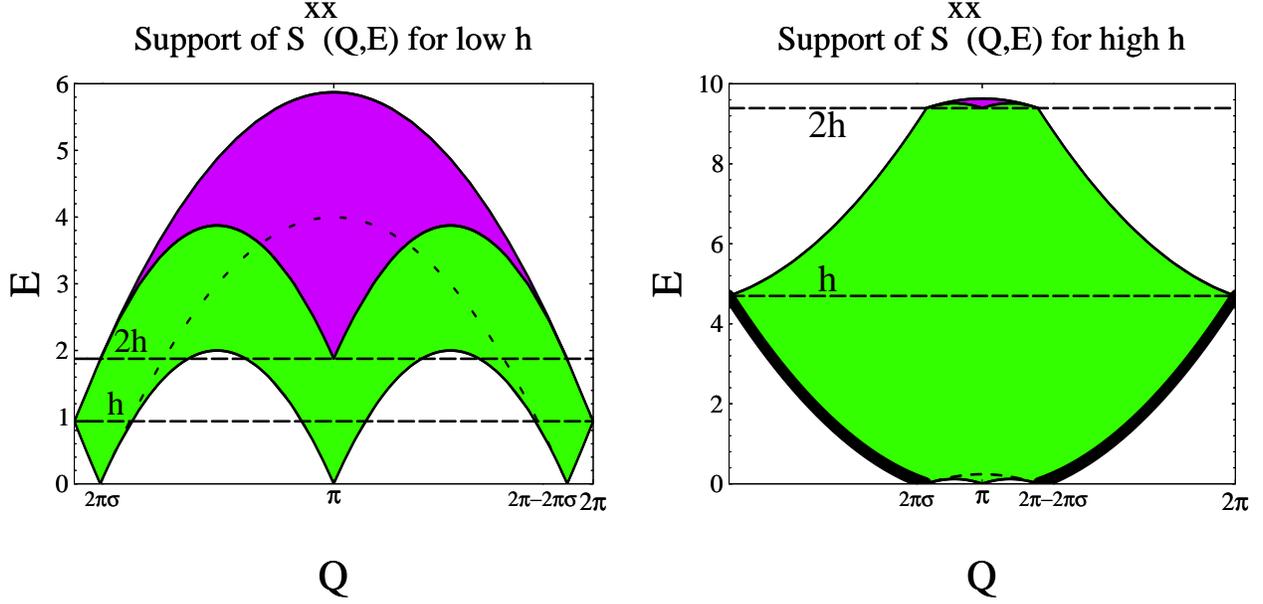}}
\vspace{.3in}
\caption{$4S^{xx}(Q,E)=S^{+-}(Q,E)+S^{-+}(Q,E)$ vs.  $(Q,E)$ for low
and high
$h$.  The contributions of $\Delta N_{sp}=+2$ (types $(iv),(v)^a$ and
$(vi)^a$)
live in the dark shaded region.  These excitations will survive in
the limit
$h\rightarrow 0$.  The $\Delta N_{sp}=0,-2$ can also occupy the area
shaded
light.  For $h\rightarrow h_c$ only the 1 magnon contributions
survive, as can
be seen in Fig.\ \protect\ref{smindata}.  These high field magnon
excitations
are indicated by the thick lines.}
\label{sminsup}
\end{figure}

\begin{figure}
\vspace{.2in}
\centerline{\epsfxsize=6.5in\epsfbox[28 382 759 699]{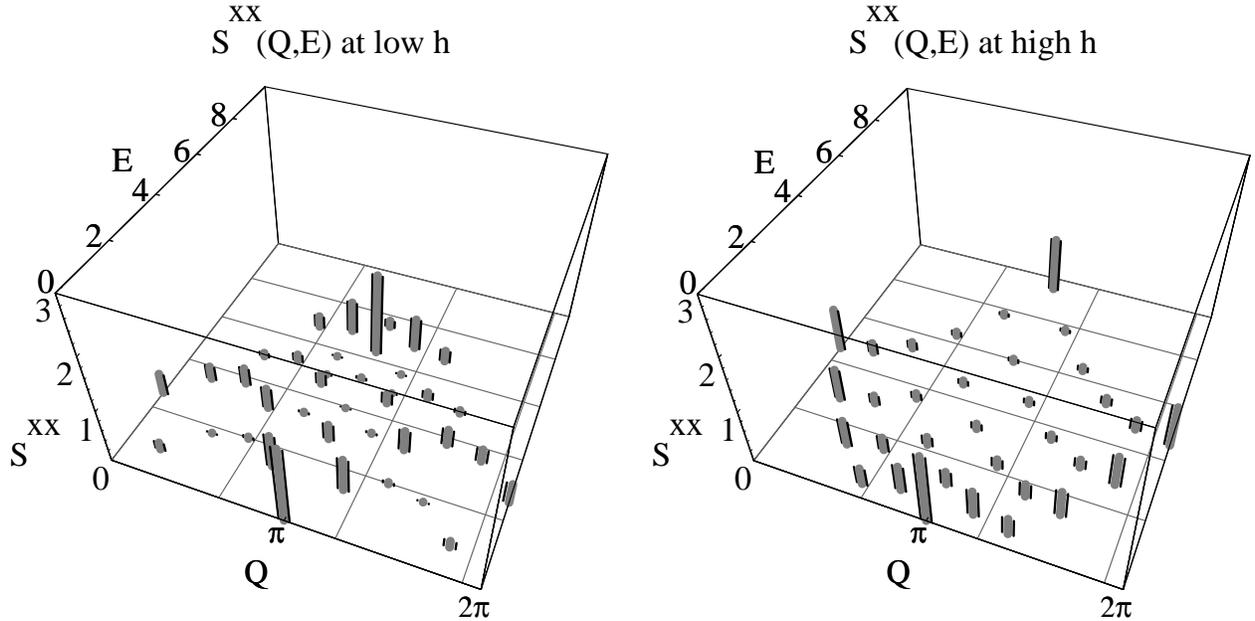}}
\vspace{.2in}
\caption{$S^{xx}(Q,E)$ for low and high $h$, from numerical
simulations on a
10-site chain. $E$ is in units of $v_s /\pi$. }
\label{smindata}
\end{figure}

Since we now know all possible excitations contributing to
$S^{+-}(Q,E)$ we can
draw its support in Fig.\ \ref{sminsup} for low and high values of
$h$.  Fig.\
\ref{smindata} show numerical data on $S^{+-}(Q,E)$ for those values
of $h$.
Table \ref{sumtable} summarizes the selection rules and available
information
on matrix elements.

\section{Comparison to the Bethe Ansatz Model}
\label{mulsec}

In 1980 M\"{u}ller et al.\ \cite{M81} did a similar calculation of
DSFs for the
nearest neighbor Heisenberg chain.  They identified certain types of
states
called {\em spin wave continuum} states (SWC) as carrying the
dominant
contribution to the DSFs.  These SWC states can be described in a
Bethe Ansatz
rapidity language by occupation sequences, just like the Yangian
multiplets in
the HSM model.  As it turns out, these SWC states correspond to {\em
exactly
the same } rapidity sequences that are favored by our selection rules
in the
HSM!  Although the dispersion relations for the BA rapidities are
different
from those in the HSM, the authors find the support of the DSFs in
the nearest
neighbor model to have essentially the same shape as we do in the the
inverse
exchange case.

However, in the nearest-neighbor Heisenberg chain there are some
``anomalous''
states, characterized by a change of more than $\pm 2$ spinons, which
contribute to a lesser degree to the structure functions and {\em
don't} lie
within the bounds found by the authors.  In the HSM these
contributions are
completely absent and once again we find this model to have a
surprisingly
clean structure.  So in this sense the Haldane-Shastry model is an
ideal spinon
gas, whereas in the NNE Heisenberg chain the spinons interact.

M\"{u}ller et al.\ also gave general rules, based on comparing
Clebsch-Gordon
coefficients, determining which matrix elements will survive in the
thermodynamic limit.  Their conclusion is that the only surviving
ones have
$S^{tot}= S^{z}$.  This means only excitations $(iii)$ for
$S^{-+}(Q,E)$ ,
$(ii)^a$ and $(ii)^b$ for $S^{zz}(Q,E)$ and $(vi)^a$, $(vi)^b$ and
$(vi)^c$ for
$S^{+-}(Q,E)$ remain relevant.  (Exceptions are single excitations
with $Q=0$,
since these correspond to $ S^+$, $ S^{z}$ and $ S^{-}$ which give
macroscopic
contributions, as we can see in the figures).  These contributions
are trivial
to compute.

\section{Conclusion}
\label{conclusion}

We found a remarkably simple selection rule for nonzero matrix
elements of
local spin operators between eigenstates of the Haldane-Shastry
model, which is
reminiscent of the ideal gas single-particle selection rules.  One of
the
consequences of this rule is that the {\em total} number of spinons
can only
change by $0,\pm 2$.  Within the occupation sequences this holds {\em
locally}
as well.

In the particular case that one of the states in the matrix element
is also the
groundstate in a magnetic field (i.e.\ fully polarized spinons,
condensed into
the left- and rightmost orbitals in equal amounts), the general
selection rule
only allows excitations with no more than 2 spinons (the rule applied
to the
center region) and one left- and one right moving magnon (the rule
applied to
the condensates on the left and right) \cite{Ja94}.  This implies
that the
structure functions based on the matrix elements involving these
states have a
finite support in a region dictated by convolving the dispersion
relations of
these particles (Figs.\ref{szzsup} and \ref{sminsup}).  These regions
have the
same shape as those for the nearest neighbor Heisenberg chain.  The
latter
model carries some weight outside these regions as well.  Therefore
the HSM
model has a much cleaner spinon structure than the BA model.

Matrix elements that connect a number of the states in these regions
to the
groundstate through the local action of a spin operator have been
presented.
However, information is lacking on those types that involve states,
not in the
spin multiplet of the YHWS.  This is particularly bothersome for
types
$(ii)^a$, $(vi)^a$ and $(vi)^b$ since these will survive in the
thermodynamic
limit.  Their calculation would allow a full reconstruction of the
$h\neq 0 $
DFS for the HSM.  A more algebraic treatment involving Yangian
operators should
provide more insight.

This work was supported in part by NSF Grant No.\ DMR922407 and an
IBM Graduate
Fellowship Award.

\acknowledgments
One of us (JCT) would like to thank Dr.  Zachary Ha for useful
discussions and
making available his preprint prior to publication.

\appendix
\section{}

We show that localized spinon wavefunctions with $N_{sp}$ spinons all
pointing
$\uparrow$ are necessarily linear combinations of YHWS with exactly
$N_{sp}$
spinons.  This follows easily from the fact that both $ S^+$ and
$J^{+}_{1}$
annihilate these states.  If we write $\Psi( n_1,\ldots,n_{M}
)=\psi(z_{n_1},\ldots,z_{n_M})$ where $z_{n} = \exp({\frac{2\pi i
n}{N}})$ and

\begin{equation}
\psi(w_1,\ldots,w_M)  =  \prod_{i<j}(w_i-w_j)^2\prod_{i}w_i
\prod_{i=1}^{M}\prod_{j=1}^{N_{sp}}(w_i-z_{\alpha_j}),
\end{equation}
then
\begin{eqnarray}
\lefteqn{( S^+\Psi)( n_{1},\ldots,n_{M-1} ) }\nonumber\\
&=&\sum_{j=1}^{N}\Psi( n_{1},\ldots,n_{M-1} ,j) = \sum_{j=1}^{N}\psi(
n_{1},\ldots,n_{M-1} ,e^{\frac{2\pi ij}{N}}) \nonumber\\
& =& \psi(z_{n_1},\ldots,z_{n_M},0) =0,
\end{eqnarray}
(where the $\sum_{j}$ was recognized as the zero-mode of a Fourier
expansion).
And with $w_{jk} = \frac{z_j + z_k}{z_j - z_k}$:
\begin{eqnarray}
&&(J_{1}^{+}\Psi)( n_{1},\ldots,n_{M-1} )\sim\nonumber\\
&&\rule{2em}{0em} -\sum_{j=1}^{M}\sum_{i\neq n_j}^{N} w_{n_j,i} \Psi(
n_{1},\ldots,n_{M-1} ,i),
\end{eqnarray}
where we used the fact that $2 S^{z}_{i} =
1-2\sum_{j=1}^{M}\delta_{i,n_j}$ and
that $\Psi$ is a symmetric function which vanishes when two of its
arguments
coincide.  We now use that when convolving a polynomial $P(z)$ of
degree less
than $N$ (such as $\psi$) with $w_{jk}$ we have:

\begin{equation}
\sum_{i=1}^{N} w_{ij}P(z_i) = N P(z_j) -2 z_j P'(z_j) - N P(0),
\end{equation}
($P'$ is the derivative of $P$).  Since $\psi$ has a double zero when
two of
its arguments coincide and it vanishes at $z=0$ we have
$J^{+}_{1}\Psi = 0$.

\section{}

In this appendix we want to prove eq.\ (\ref{spin2exp}).  Let us
first
introduce the following identity which holds for any set of distinct
complex
numbers $\{ \omega_i\}$:

\begin{equation}
\prod_{j=1}^{M-1}(Z_j-z)=\sum_{k=1}^{M}\prod_{l(\neq
k)}^{M}\frac{\omega_l
-z}{\omega_l - \omega_k}\prod_{i=1}^{M}(Z_i -\omega_k).
\label{lagrid}
\end{equation}
The RHS is just the Lagrange interpolation formula applied to
the function in $z$ on the LHS!

Say we want to write $S_{i}^{+}\Psi_{\{\alpha_i\} }$ as a linear
combination of
localized spinon wavefunctions with two more spinons than
$\Psi_{\{\alpha_i\}
}$, and all spinons pointing $\uparrow$.  We fix one of the
additional two
spinons at $i$, the site on which $S^{+}_{i}$ acts, i.e.\ :

\begin{eqnarray}
\lefteqn{( S^{+}_{i}\Psi_{\{\alpha_i\} })( n_{1},\ldots,n_{M-1}
)}\nonumber\\
&\equiv &\Psi( n_{1},\ldots,n_{M-1}
,i|\alpha_1,\ldots,\alpha_{N_{sp}})
\nonumber\\
&=&\sum_{p(\neq i,\{\alpha_k\})}a_p\Psi( n_{1},\ldots,n_{M-1}
|\alpha_1,\ldots,\alpha_{N_{sp}},i,p).
\end{eqnarray}
Using eq.\ (\ref{locspinwf}) we can divide out common factors of
$(z_{n_k}-z_{n_l})$ etc.\ , and we are left with:

\begin{equation}
z_i \prod_{j}(z_{n_j}-z_i) = \sum_{p\neq (i,\{\alpha_k\}
)}a_p\prod_{j}(z_{n_j}-z_p).
\end{equation}
The result follows when we apply the identity (\ref{lagrid}) to this
equation
with $z=z_{i}$ and $a_p=z_{i}\prod_{k}\frac{z_k - z_i}{z_k -z_p}$
where the
$z_k$ are randomly chosen distinct sites which don't coincide with
the
localized spinons.

\section{}

We show that $ S^{z}_{i}\left|{\rm GS}\right.  \rangle$ can only have
0 or 2
more spinons than $\left|{\rm GS}\right.  \rangle$ where $\left|{\rm
GS}\right.
\rangle$ is a YHWS groundstate in a nonzero magnetic field with

\begin{eqnarray}
\langle\left.  n_1,\ldots,n_{M} \right|{\rm GS}\rangle &=&\Psi_{0}(
n_1,\ldots,n_{M} ) \nonumber\\
&=& \prod_{i<j}(z_{n_i}-z_{n_j})^2\prod_{i}z_{n_i}^{\frac{N}{2}-M+1}.
\end{eqnarray}
The proof hinges on the fact that in either of those cases ($\Delta
N_{sp} =0$
or $\Delta N_{sp} =+2$) acting on $ S^{z}_{i}\left|{\rm GS}\right.
\rangle$
{\em twice} with $J_{1}^{+}$ will annihilate that state.  Potential
$\Delta
N_{sp} =4,\ldots$ contributions should survive as they are at least 2
levels
from the top of their Yangian multiplet.  Now

\begin{equation}
(2 S^{z}_{i} \Psi_{0})\left({\scriptstyle n_1,\ldots,n_{M} }\right) =
\left(1-2\sum_{j=1}^{M} \delta_{i,n_j}\right)
\Psi_{0}\left({\scriptstyle
n_1,\ldots,n_{M} }\right),
\end{equation}
and

\begin{eqnarray}
&&\left( 2J_{1}^{+} S^{z}_{i}\Psi_{0}\right) ( n_{1},\ldots,n_{M-1}
)\sim\nonumber\\
&&\rule{1em}{0em} - \sum_{k=1}^{M-1}\sum_{l\neq n_k}^{N} w_{n_k,l}
\times
\nonumber\\
&&\rule{2em}{0em} \left[\Psi_{0}({\scriptstyle n_{1},\ldots,n_{M-1}
,l})-2\sum_{j=1}^{M-1}\delta_{i,n_j}\Psi_{0}({\scriptstyle
n_{1},\ldots,n_{M-1}
,l})\right] \nonumber\\
& &\rule{1em}{0em}-2\sum_{k=1}^{M-1}w_{n_k,i}\Psi_{0}(
n_{1},\ldots,n_{M-1}
,i).
\end{eqnarray}
The first two terms vanish when we apply the convolution theorem with
$w_{kl}$,
as in Appendix A; only the last term survives.  Notice that the
$\{n_i\};i=1\ldots,M-1$ cannot be equal to $i$ anymore.  This state
is
trivially annihilated by $ S^+$ as it vanishes at $z=0$.
Furthermore:

\begin{eqnarray}
&&\left(\left(J_{1}^{+}\right)^2 S^{z}_{i}\Psi_{0}\right)(
n_{1},\ldots,n_{M-2}
) = \nonumber\\
& &\rule{1em}{0em} \sum_{p=1}^{M-2}\sum_{q\neq i,n_p}^{N}
w_{n_p,q}\sum_{k=1}^{M-2} w_{n_k,i}\Psi_{0}( n_{1},\ldots,n_{M-2}
,q,i)
\nonumber\\
& &\rule{1em}{0em} + \sum_{p=1}^{M-2}\sum_{q\neq i,n_p}^{N} w_{n_p,q}
w_{q,i}\Psi_{0}( n_{1},\ldots,n_{M-2} ,q,i).
\end{eqnarray}
With the help of the convolution theorem we can set the first term to
zero (we
can stick in the extra term with $q=i$, which is missing, at no cost
since
$\Psi_{0}$ vanishes when $q=i$).  For the second term we use the
identity that
lies at the heart of the integrability of the Haldane-Shastry type
models:
$w_{ij}w_{jk}+w_{jk}w_{ki}+w_{ki}w_{ij} = -1$.

\begin{eqnarray}
\lefteqn{\left(\left(J_{1}^{+}\right)^2 S^{z}_{i}\Psi\right)(
n_{1},\ldots,n_{M-2} ) =} \\
& &-\sum_{p=1}^{M-2}\sum_{q\neq
i,n_p}^{N}(1+w_{i,n_p}w_{n_p,q}+w_{q,i}w_{i,n_p})\Psi({\scriptstyle
n_{1},\ldots,n_{M-2} ,q,i})\nonumber.
\end{eqnarray}
The first term is zero since $\Psi$ is a polynomial that vanishes at
the
origin, whereas terms two and three can also be put to zero with the
help of
the convolution theorem.

\widetext
\begin{table}
\caption{List of matrix elements contributing to the DSF.  The
correlation
functions in the last column refer to those in the
Calogero-Sutherland model as
they appeared in the previous sections.  in eqs.\
(\protect\ref{psidagpsi},
\protect\ref{psidagpsisz}, \protect\ref{rhorho},
\protect\ref{psidagpsismin},
\protect\ref{rhorhosmin}, \protect\ref{psipsidag}).  The entries
marked with a
$(\dag )$ do not survive in the thermodynamic limit.  For the example
occupation sequences we act on a groundstate
$000\protect\rule[-.25em]{.3ex}{1.1em}
101010101\protect\rule[-.25em]{.3ex}{1.1em} 000$, with the
$\protect\rule[-.25em]{.3ex}{1.1em}$ delimiting the center region.
In the
column under ``Excitation'' ${\cal S}$ denotes a spinon and ${\cal
M}$ denotes
a magnon.  }

\samepage
\begin{tabular}{ccccccc}
\nopagebreak
DSF & Type & $S^{tot}$ & $\Delta N_{sp} $ &Typical contributing &
Excitation &
Matrix element \\
    &      &           &            & Yangian multiplet   &
 &    \\
\tableline
$S^{-+}(Q,E)$ & $(iii)$ &$S=S_0 +1$& $\Delta N_{sp} =+2$ &
$000\rule[-.25em]{.3ex}{1.1em} 100100101\rule[-.25em]{.3ex}{1.1em}
000$ &
$2{\cal{S}}$ &$\langle\Psi^{\dag}\Psi\rangle$ \\
\nopagebreak
\rule{2em}{0em}${\scriptstyle ( S^{z} = S_0 +1)}$& & & & & &\\
\nopagebreak
& & & & & &\\
\nopagebreak
$S^{zz}(Q,E)$ &$(i)^{\dag}$ & $S=S_0 +1$ & $\Delta N_{sp} =+2$
&$000\rule[-.25em]{.3ex}{1.1em} 100101001\rule[-.25em]{.3ex}{1.1em}
000$ &
$2{\cal{S}}$ & $\frac{1}{2( S_{0} +1)}\langle\Psi^{\dag}\Psi\rangle$
\\
\nopagebreak
\rule{2em}{0em}${\scriptstyle ( S^{z} = S_0)}$ & $(ii)^{a\dag}$&
$S=S_0 $ &
$\Delta N_{sp} =+2$ &$000\rule[-.25em]{.3ex}{1.1em}
100101001\rule[-.25em]{.3ex}{1.1em} 000$ & $2{\cal{S}}$ & \\
\nopagebreak
&$(ii)^b$ & $S=S_0$ & $\Delta N_{sp} =0$ &
$100\rule[-.25em]{.3ex}{1.1em}
101001001\rule[-.25em]{.3ex}{1.1em} 000$ & $2{\cal{S}} + {\cal{M}}$ &
$\langle\rho\rho\rangle$ \\
\nopagebreak
& & & & & &\\
\nopagebreak
$S^{+-}(Q,E) $ & $(iv)^{\dag}$& $S=S_0 +1$ & $\Delta N_{sp} =+2$
&$000\rule[-.25em]{.3ex}{1.1em} 100100101\rule[-.25em]{.3ex}{1.1em}
000$ & $
2{\cal{S}}$ & $\frac{1}{( S_{0} +1)(2 S_{0} +
1)}\langle\Psi^{\dag}\Psi\rangle
$ \\
\nopagebreak
\rule{2em}{0em}${\scriptstyle ( S^{z} = S_0 -1)}$ & $(v)^{a\dag}$&
$S=S_0 $ &
$\Delta N_{sp} =+2$ & $000\rule[-.25em]{.3ex}{1.1em}
010010101\rule[-.25em]{.3ex}{1.1em} 000$ & $2{\cal{S}}$ & \\
\nopagebreak
& $(vi)^a$& $S=S_0 -1$ & $\Delta N_{sp} =+2$
&$000\rule[-.25em]{.3ex}{1.1em}
010010101\rule[-.25em]{.3ex}{1.1em} 000$ & $2{\cal{S}}$ & \\
\nopagebreak
& $(v)^{b\dag}$& $S=S_0 $ & $\Delta N_{sp} =0$ &
$000\rule[-.25em]{.3ex}{1.1em}
100100101\rule[-.25em]{.3ex}{1.1em} 010$ & $2{\cal{S}} + {\cal{M}}$ &
$\frac{2}{ S_{0}}\langle\rho\rho\rangle $ \\
\nopagebreak
& $(vi)^b$& $S=S_0 -1$ & $\Delta N_{sp} =0$ &
$100\rule[-.25em]{.3ex}{1.1em}
100101001\rule[-.25em]{.3ex}{1.1em} 000$& $2{\cal{S}} + {\cal{M}}$ &
\\
\nopagebreak
& $(vi)^c$& $S=S_0 -1$ & $\Delta N_{sp} =-2$ &
$100\rule[-.25em]{.3ex}{1.1em}
100100101\rule[-.25em]{.3ex}{1.1em} 010$ & $2{\cal{S}} +
2{\cal{M}}$, ${\cal{M}}$ & $\langle\Psi\Psi^{\dag}\rangle$\\
\nopagebreak
\end{tabular}
\label{sumtable}
\end{table}
\end{document}